\documentstyle[epsfig]{elsart}
\def\be{\begin{eqnarray}}
\def\ee{\end{eqnarray}}
\newcommand{\ep}{\varepsilon}
\newcommand{\omnu}{\omega_{\nu}}

\newcommand{\bp}{\vec{p}}
\newcommand{\bq}{\vec{q}}

\newcommand{\sla}{\! \not \!}
\newcommand{\anu}{\bar\nu}

\begin{document}
\begin{frontmatter}
\title{Coherence Effects and Neutrino Pair Bremsstrahlung 
      in Neutron Stars}

\author{Armen Sedrakian and Alex Dieperink}
\address{Kernfysisch Versneller Instituut,
         NL-9747 AA Groningen,
         The Netherlands
         }

\maketitle
\begin{abstract}
\noindent
We calculate the rate of energy radiation by bremsstrahlung
of neutrino pairs by baryons in neutron stars employing 
a transport model where neutrinos couple to baryons 
with spectral width. The coherence effects, which are
included by computing the self energies with fully dressed 
propagators, lead to the Landau-Pomeranchuk-Migdal suppression 
of the neutrino emission rates in the soft neutrino and high 
temperature limit. A microscopic evaluation of 
the bremsstrahlung  rate by neutrons is carried out using the 
Brueckner theory of nuclear matter at finite temperatures.
\end{abstract}
\end{frontmatter}

\section{Introduction}

Neutrino pair bremsstrahlung by baryons, $B_1+B_2\to B_1+B_2 
+\nu+\overline\nu$, is among the dominant processes by which 
neutron stars radiate away their energy during the first 
several thousand years of their evolution.  
The rates of these processes have been derived
by Friman and Maxwell ~\cite{FRIMAN_MAXWELL} and 
Voskresensky and Senatorov ~\cite{VOSKRESENSKY1} employing
the Fermi-liquid interaction combined with 
different prescriptions for pion exchange contribution to the 
nuclear matrix element. 

The spectator in the reaction above
is needed to insure that the process is allowed 
kinematically.\footnote{The neutrino 
pair bremsstrahlung via the reaction $B\to B +\nu+\overline\nu$ is
inefficient as the energy conservation requires $\omega \sim q\, v_F$ where 
$q$ is the momentum  transfer and $v_F$ is the baryon Fermi velocity, 
while  for neutrinos in the time-like region $\omega\ge qc$.  
In more formal terms  the imaginary part of the 
quasi-particle single-loop polarization function vanishes in the 
low-temperature limit.}  If the formation length of
the neutrino radiation is of the order of the mean
free path of a baryon, the role of the spectator is taken 
over by the medium because the baryon undergoes multiple 
scattering during the radiation. Then the reaction 
rate is subject to  the 
Landau-Pomeranchuk-Migdal (LPM) suppression well-know from QED.
At not too low temperatures the LPM suppression of the 
brems\-strahlung could be a factor in neutron stars,
for the radiation frequency is of the 
order of temperature and could have a 
magnitude comparable to the baryon's collisional width.

The LPM effect has been naturally recovered in  quantum 
many-body theory in attempts to overcome the 
limitations of the quasi-particle approximation 
which led to infrared-divergent results in the soft limit.
Knoll and  Voskresensky, ref. \cite{KNOLL_VOSKRESENSKY}, 
developed  a general approach to the LPM effect using
the Schwinger-Keldysh formulation of 
non-equilibrium  field theory.
The particle production rates in their approach are calculated 
from an expansion of the self-energies in terms of 
skeleton (or compact) diagrams with fully dressed 
particle propagators. Raffelt
and Hannestad and Raffelt~\cite{RAFFELT} 
have discussed in the supernova context 
the neutrino pair bremsstrahlung using the one-pion 
exchange interaction and including leading order 
corrections in particle width.

In this Letter  we shall adopt the formalism of 
ref. \cite{KNOLL_VOSKRESENSKY} to compute the rate of 
the neutrino pair bremsstrahlung from baryonic matter 
in neutron stars and supernovae, including the LPM effect.
The spectral width of baryon propagators, which enters the 
rates of the bremsstrahlung, is derived from the finite 
temperature Brueckner theory of nuclear matter. 
Friman and Maxwell's \cite{FRIMAN_MAXWELL}
result is adopted as the reference for comparison.

The rates of neutrino pair bremsstrahlung in the Schwinger-Keldysh 
technique can be derived using the formalism of the 
optical theorem \cite{VOSKRESENSKY2}.
Here we follow a somewhat different path. It is known
that the rates of the neutrino emission can be 
recovered from the Boltzmann equation for neutrinos \cite{IWAMOTO}.
For present purposes the Boltzmann equation, which deals only with  
on-shell particles, is not sufficient. We shall
recover neutrino emission rates 
form the Kadanoff-Baym transport equation \cite{KADANOFF_BAYM}
which includes the collisional coupling of neutrinos to baryons 
with a finite spectral width. As the dense cores of neutron star 
develop, apart from neutrons and protons, a large 
equilibrium fraction of hyperons~\cite{GLENDENNING},
our discussion shall refer generally to baryons. We treat
neutrinos  as relativistic massless 
Dirac fermions within the relativistic Kadanoff-Baym 
transport theory \cite{BOTERMANS_MALFLIET,HEINZ}.
Our final expression relating the neutrino 
emissivity to the polarization 
of the medium agrees with the one derived
by Voskresensky and Senatorov \cite{VOSKRESENSKY2} from 
the optical theorem. 

\section{Formalism}

We start from the Kadanoff-Baym transport equation  for the 
neutrinos:\footnote{Terms which do not contribute to the 
left-hand-side of the transport equation in the mean field 
approximation are dropped.}
\be\label{BE_NU}
 \{\gamma^{\mu}, \partial_{x\mu} S^<(q,x) \}_{+}  =
 \sigma^<(q,x)S^>(q,x)
-\sigma^>(q,x)S^<(q,x),\nonumber\\
\ee
where $q\equiv (\vec q , q_0)$ and  $x$ are the neutrino four momentum 
and the center-of-mass space-time coordinate, respectively,
$S^{>,<}$ and $\sigma^{>,<}(q,x)$ are the neutrino propagators and 
collisional self-energies, which are real-time contour
ordered with fixed-time arguments on the lower (upper) and
upper (lower) branches of the Schwinger-Keldysh contour,
respectively.
Treating neutrinos on-mass-shell, the propagators can 
be expressed in terms of the non-equilibrium distribution functions
(Kadanoff-Baym ansatz \cite{KADANOFF_BAYM})  
\be\label{PROP}
S^<(q,x)
= \frac{i\pi\sla q}{\omnu( q)}
\Big[ \delta\left(q_0-\omnu(\bq)\right)f_{\nu}(q, x)
-\delta\left(q_0+\omnu(\bq)\right) \left(1-f_{\bar \nu}
(-q,x)\right) \Big];\nonumber \\
\ee
the $S^>(q,x)$ propagator follows from eq. (\ref{PROP})
via the interchange $f_{\nu}\leftrightarrow (1-f_{\nu})$. 
The collision integrals contain neutrino
scattering, absorption and emission contributions of which only
the latter one is relevant for the cold neutron stars.

The (anti-)neutrino emissivity (the power of the
energy radiated per volume unit) is obtained by multiplying the
left-hand-side of the transport equation by the neutrino energy
and integrating over an element
of the four dimensional phase-space:
\be\label{E_NUNU_PRE}
\epsilon_{\nu\bar\nu}=\frac{d}{dt}\sum_f\int\!\frac{d^3q}{(2\pi)^3}
&&\left[f_{\nu}(\bq) +f_{\bar\nu}(\bq)\right]\omnu(\bq)\nonumber \\
&&\hspace{1cm} =\sum_f
\int\!\frac{d^3q}{(2\pi)^3}\left[I_{\nu}^{\rm <, em}(\vec q)
-I_{\anu}^{\rm >, em}(\vec q)\right]\omnu(\bq),
\ee
where the reduced collision integral $I_{\nu}^{\rm <, em}( q)$
(em stands for emission) originates from energy integration
over the range $[0,\infty]$ and corresponds to the neutrino
branch of the spectrum, while 
$I_{\anu}^{\rm >, em}( q)$ results from the integration
over the range $[-\infty, 0]$ and corresponds to the anti-neutrino
branch of the spectrum; the sum is over 
the neutrino flavors. 
The leading order Feynman diagrams in the expansion of the 
self-energies $\sigma^{>,<}$ with respect to the 
weak neutrino-baryon coupling are shown in the Fig. 1a.
\begin{figure}
\begin{center}
\mbox{\epsfig{figure=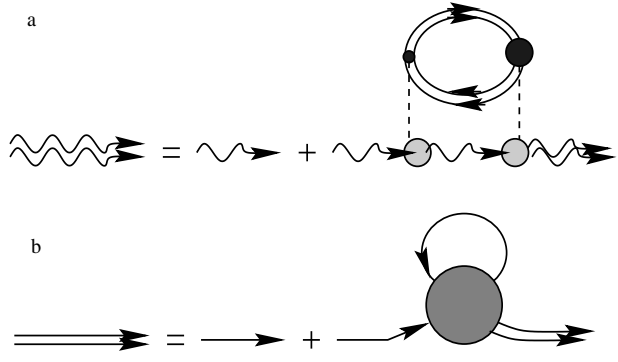,height=1.8in,width=3.2in,angle=0}}
\end{center}
\caption[]
{\footnotesize{Dyson equation for (a) neutrinos (wavy lines)  
and (b) baryons (solid lines). Double and single lines correspond 
to dressed and bare propagators, respectively, the dashed lines
correspond  to the $W,Z$ boson exchange. The light dots 
denote the neutrino weak interaction vertex. The  
dark dot stands for the particle-hole vertex renormalization
in the baryon loop. 
The shaded block in the lower panel is the thermodynamic $T$-matrix 
in the lowest order Brueckner theory.}}
\label{fig1}
\end{figure} 
The corresponding one-loop transport self-energies are read-off from
the diagram
\be\label{SIGMA}
&&-i\sigma^{>,<}(q_1,x) = \int \frac{d^4 q}{(2\pi)^4}
\int \frac{d^4 q_2}{(2\pi)^4}(2\pi)^4 \delta^4(q_1 - q_2 - q)\nonumber \\
&&\hspace{5.5cm}i\Gamma_{q}^{\mu}\, iS^{>,<}(q_2,x) 
i\Gamma_{q}^{\dagger\,\lambda}i\Pi_{\mu\lambda}^{>,<}(-q,x),
\ee
where $\Pi_{\mu\lambda}^{>,<}(q)$ are the Fourier transforms of the
baryon polarization functions with time arguments fixed on the
upper and lower branches of the contour. The baryon propagators in 
$\Pi_{\mu\lambda}^{>,<}(q)$ are fully dressed; the Fig. 1b displays 
the baryon propagator dressing due to the strong interaction
within the lowest order Brueckner theory used 
in the numerical evaluation below. For small energy-momentum 
transfers involved in the problem, the weak interaction 
vertex, $\Gamma_q^{\mu}$, can be replaced by the contact interaction:
$
\Gamma^{\mu} = \left(G/2\sqrt{2}\right) 
\gamma^{\mu}(1-\gamma^{5})
$, where $G$ is the weak coupling constant.

Inserting the self-energies and the propagators in the 
left-hand-side of the transport equation (\ref{BE_NU}) 
for the neutrino branch, we find, e.g., for the first 
term which is the gain part of the collision integral
\be\label{GAIN}
I_{\nu}^{\rm <}( q_1,x)&=& -i\int_{0}^\infty \frac{dq_{10}}{2\pi}
 {\rm Tr}
\Biggl\{ \int \frac{d^4 q}{(2\pi)^4}
\int \frac{d^4 q_2}{(2\pi)^4}(2\pi)^4 \delta^4(q_1 - q_2 - q)
\Gamma^{\mu}\frac{\pi\sla q_2}{\omnu( q_2)}\nonumber \\
&&
\Big[\delta\left(q_{02}-\omnu( \bq_2)\right)f_{\nu}(q_2, x)
-\delta\left(q_{02}+\omnu(\bq_2)\right) \left(1-f_{\bar \nu}
(-q_2,x)\right) \Big]\nonumber\\
&& \Gamma^{\dagger\,\lambda} \frac{\pi\sla q_1}{\omnu( q_1)}
\delta\left(q_{10}-\omnu(\bq_1)\right)\left(1-f_{\nu}(q_1,x)\right)
\Pi_{\mu\lambda}^{>}(-q,x)\Biggr\}.
\ee
One may identify various contributions to the collision integral: the 
terms $\propto (1-f_{\nu}) f_{\nu}$ and 
$\propto (1- f_{\nu})(1-f_{\anu})$
in the gain integral correspond to the scattering-in
and emission processes, respectively.
Similarly the loss term, $I^{\rm >}_{\nu}$,
which is obtained from eq. (\ref{GAIN}) via the simultaneous
interchange $f_{\nu,\anu}\leftrightarrow (1-f_{\nu,\anu})$ and
replacement of $\Pi_{\mu\lambda}^>$ by $\Pi_{\mu\lambda}^<$, 
contains the terms
$\propto f_{\nu,\anu}(1-f_{\nu})$ and $\propto f_{\nu}f_{\anu} $
corresponding to  scattering-out and absorption, respectively.
As well known, e.g. ref. \cite{RAFFELT},
the neutrino mean free path exceeds the stellar radius and 
neutrinos leave the star without interactions
except for several seconds after the neutron star birth. If 
neutrinos are untrapped, the only terms that contribute  
to the collision integral are the  $\nu\anu$  
emission terms, one present in the neutrino gain integral (\ref{GAIN}) 
and the other in the anti-neutrino loss integral, $I_{\anu}^{\rm >}$ 
(which follows from eq. (\ref{GAIN}) by replacing $\Pi^>_{\mu\lambda}$ 
by $\Pi^<_{\mu\lambda}$ and 
changing the integration limits).

Next we enforce the quasi-particle approximation in the collision 
integrals  for neutrinos and 
anti-neutrinos by carrying out the  $q_{02}$ and $q_{01}$ energy 
integrations. The gain and loss terms of the collision integrals
are combined using the identities $\Pi_{\mu\lambda}^{<}(q)
=\Pi_{\mu\lambda}^{>}(-q) = 2i g_B(q_0) {\rm Im}\,
\Pi_{\mu\lambda}^R(q)$, where $g_B(q_0)$ is the Bose function
and $\Pi^R_{\mu\lambda}(q)$ is the retarded component 
of the polarization function.
Substituting the electroweak vertex in eq. 
(\ref{E_NUNU_PRE}) and dropping the spatial argument, we find 
\be\label{EMISSIVITY}
\epsilon_{\nu\anu}&=& - 2\left( \frac{G}{2\sqrt{2}}\right)^2
\sum_f\int\!\frac{d^3q_2}{(2\pi)^32 \omnu( q_2)}
\int\!\frac{d^3 q_1}{(2\pi)^3 2\omnu( q_1)}
\int\!
\frac{d^4 q}{(2\pi)^4}
\nonumber\\
&&\hspace{1cm}
(2\pi)^4 \delta^3(\bq_2 + \bq_1 -  \bq)\delta(\omnu(\bq_2)+\omnu(\bq_1)-q_{0})
 \nonumber\\
&&\hspace{2cm}
\left[\omnu(\bq_2)+\omnu(\bq_1)\right]\, g_B(q_0) 
 \Lambda^{\mu\lambda}(q_1,q_2){\rm Im}\,\Pi_{\mu\lambda}^R(q),
\ee
where $\Lambda^{\mu\lambda}(q_1,q_2) ={\rm Tr}[
\gamma^{\mu}(1 - \gamma^5)  \sla q_1
\gamma^{\lambda}(1 - \gamma^5) \sla q_2]$ 
is the trace of the neutrino currents.
Here we made the approximation $f_{\nu} , f_{\anu} \ll 1$ 
appropriate for untrapped neutrinos.
The final expression for the emissivity, eq. (\ref{EMISSIVITY}),
recovered here from the transport equation, agrees with that 
derived from the non-equilibrium $S$-matrix 
expansion~\cite{VOSKRESENSKY2}. Analogous expressions 
for the scattering and absorption in equilibrium have been 
derived recently in refs. \cite{REDDY,BORROWS}.

The Lorentz structure of the 
baryon polarization function $\Pi_{\mu\lambda}(q)$ 
is given by the trace of the baryon transition currents 
 $$X_{\mu\lambda}(p,q) ={\rm Tr}[ 
\gamma_{\mu}(c_V - c_A\gamma_5)(\sla p +m) 
\gamma_{\lambda}(c_V - c_A\gamma_5)(\sla p +\sla q +m)] .$$
The contraction of the neutrino trace with the non-relativistic 
reduction of $X_{\mu\lambda}$ simplifies the factor
$\Lambda^{\mu\lambda}(q_1,q_2){\rm Im}\,
\Pi_{\mu\lambda}^R(q) =8\omega_{\nu}(\vec q_1)\omega_{\nu}(\vec q_2)
[c_V^2 L^R_V(\omega, \bq)  
+3c_A^2 L^R_A(\omega, \bq)]$ where the vector ($L_V^R$) and the 
axial vector ($L_A^R$)  one-loop polarization functions 
differ by the particle-hole vertex renormalization. 
The vector part corresponds to the scalar particle-hole 
vertex; the axial-vector part corresponds to the 
spin exchange interaction vertex. These renormalizations can 
be carried out \cite{VOSKRESENSKY2} in the Fermi-liquid theory, 
using the quasi-particle approximation, and
amount to replacing the weak-coupling parameters $c_V$ 
and $c_A$ with the effective ones (explicit expressions are given 
in ref. \cite{VOSKRESENSKY2}). Eventually one is left with 
the factor $8\omega_{\nu}(\vec q_1)\omega_{\nu}(\vec q_2)
(c_V^2  +3c_A^2) L^R_0(\omega, \bq)$ where 
\be\label{PI}
L_{0}^R(\omega, \bq) = 2\int\!\!\frac{d^3pd\ep'd\ep}{(2\pi)^5}
\left\{\frac{g^>(\bp,\ep')  g^<( \bp+ \bq,\ep)
-g^<( \bp,\ep') g^>( \bp + \bq,\ep)}{\ep'
-\ep+\omega+i\delta} \right\},\nonumber\\
\ee
which is the driving term (and hence the subscript 0)
in the resummation series for the particle-hole vertex;
the parameters $c_V$ and $c_A$ here and below should be 
understood as the renormalized ones.
The baryon propagators, $g^{>,<}(p)$, are related to the spectral 
function $a(p)$ 
and the Fermi distributions $f_N(p)$ via the Kadanoff-Baym
ansatz for baryons~\cite{KADANOFF_BAYM}:
$ -ig^{<}(p)= a(p) \, f_N(p)$, $g^<(p)-g^>(p) = ia(p)$,
which is exact in the equilibrium limit. The spectral function, 
in turn, is given by 
\be\label{SF} 
a(\omega,\vec p)=-2 {\rm Im}\left[ \omega -\ep_p 
+i\gamma(\omega, \vec p)/2\right]^{-1},
\ee
where the quasi-particle energy $\varepsilon_p = \varepsilon_p^0  + 
{\rm Re}\Sigma^R(\omega, \vec p)|_{\omega = \ep_p}$, 
$\varepsilon_p^0 \equiv p^2/2m$ 
(i.e. the pole of the spectral function) and the inverse lifetime
$\gamma(\omega,\vec p) = -2{\rm Im} \Sigma^R(\omega,\vec p)$ 
(i.e. the width of the spectral function) 
are given in terms of the retarded baryon self-energy, $\Sigma^R(p)$;
our approximation for the latter quantity is depicted in Fig. 1b.
We note in passing that the imaginary part of the  
quasi-particle polarization function (the limit $\gamma(p) \to 0$)
vanishes in the time-like region, $q_0^2\ge {\vec q}^2$, 
in the low temperature limit. Eq. (\ref{PI}), therefore, 
is the leading order compact diagram contributing
to the neutrino emission rate.

\section{Results}
\begin{figure}
\begin{center}
\mbox{\epsfig{figure=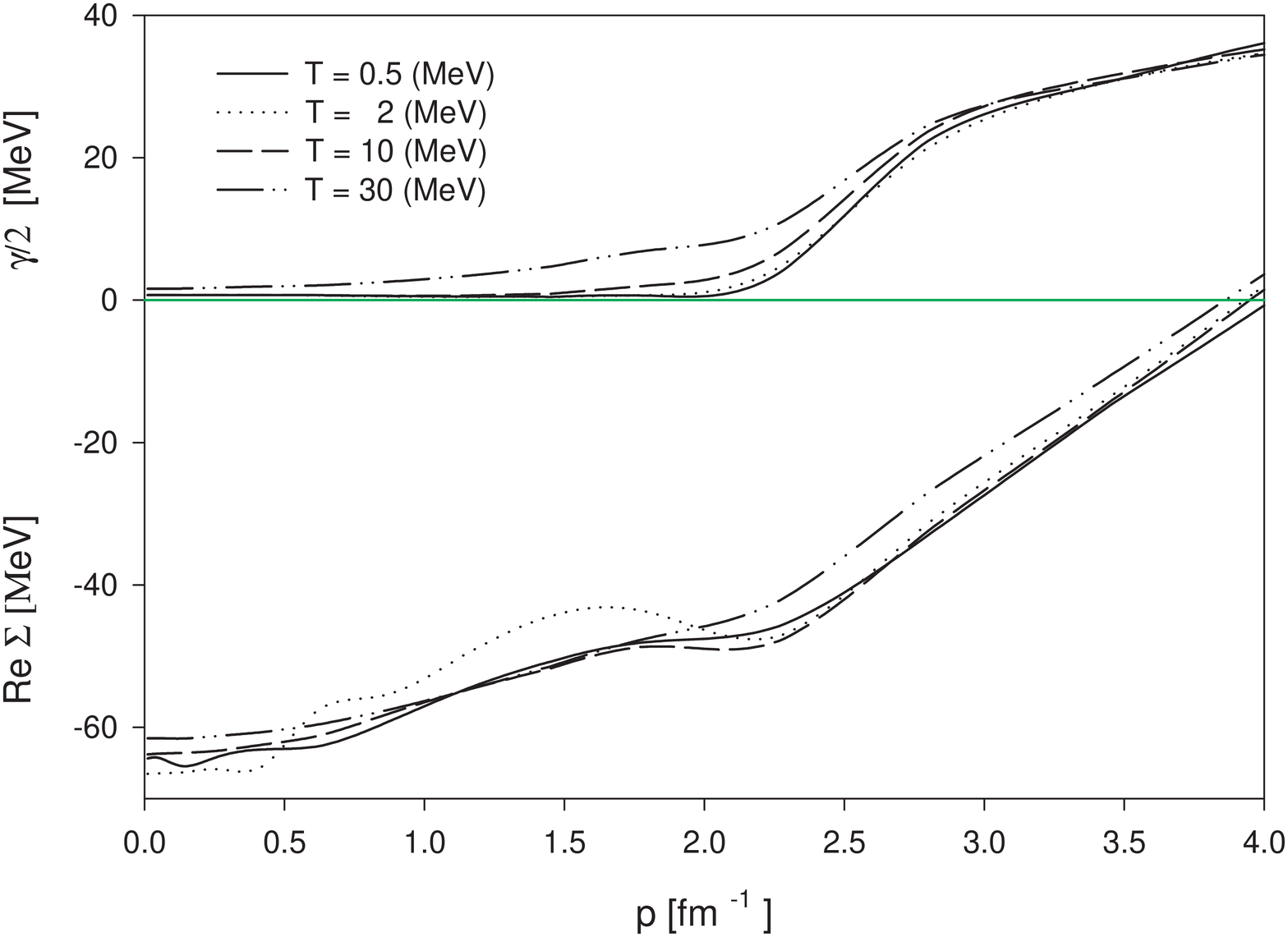,height=6.in,width=5.in,angle=0}}
\end{center}
\caption[]
{\footnotesize{The real part of the on-shell 
self-energy and the half-width 
as a function of particle momentum at the saturation density $n_s$
for different temperatures; the zero temperature
Fermi momentum is 1.7 fm $^{-1}$.}}
\label{fig2}
\end{figure}
Next we proceed to microscopic calculation of the 
self-energy and polarization function 
for the neutron matter, which is the dominant constituent of the 
dense stellar matter for densities $n \le 2 n_s$  ($n_s = 0.16$ 
fm$^{-3}$). The single particle energies and the width have
been evaluated in the self-consistent lowest order
Brueckner scheme at finite temperatures (ref. 
\cite{ALM} and references therein).
This scheme requires a self-consistent 
solution of the coupled equations for the 
thermodynamic (retarded) $T^R$-matrix
\be\label{TMAT}
T^{R\alpha}_{ll'}( p,  p', P, \omega)  &=& 
V^{\alpha}_{ll'}(p, p') \nonumber\\
&+&\frac{2}{\pi}\sum_{l''}
\int\!\! dp''\, p''^2 \, V^{\alpha}_{ll''}(p, p'')
G_2^R(p'', P, \omega)
T^{R\alpha}_{l''l'}(p'', p', P,\omega), 
\ee
where $\alpha$ collectively denotes the quantum 
numbers $(S,J,M)$ in a given partial wave, $p$ and $P$
are the magnitudes of the relative and total momentum
respectively,  $V(p,p')$ 
is the bare nuclear interaction, $G_2^R$ is the 
angle averaged two-particle propagator
\be 
G_2^R(p, P, \omega) = \int\!\frac{d\Omega}{4\pi}
\frac{\left[1-f_N(\ep(\vec P/2+\vec p))\right]
\left[1-f_N(\ep(\vec P/2-\vec p))\right]}
{\omega -\ep(\vec P/2+\vec p)-\ep(\vec P/2-\vec p)+i\delta },
\ee
with $\ep(p) = \ep_p^0 + {\rm Re}\Sigma(\ep_p,\vec p)$
and the retarded self-energy
\be\label{SELF_E} 
\Sigma^R(p, \omega) = \frac{1}{\pi}\sum_{l\alpha}(2J+1)
\int\! dp'\, p'^2 T^{R\alpha}_{ll}(p,p';p,p';\omega+ \ep(p'))
f_N(\ep(p'))~; \nonumber \\
\ee 
here the lower case $p$ is the  
magnitude of the proper momentum of a particle. 
The coupled equations (\ref{TMAT}) and (\ref{SELF_E})
are subject to normalization to the total density at a given
temperature.
Fig. 2 displays the real part of the on-shell self energy and the
half width of the spectral function as a function of the particle 
momentum for several values of the temperature at the 
saturation density $n_s$. The numerical calculations 
were carried out using the
Paris $NN$ intercation keeping partial waves 
with $L\le 2$.

Let us next turn to the imaginary part of the 
polarization function, eq. (\ref{PI}).
One of the integrations
can be removed by the delta function; further only the  
angular integral can be performed 
analytically ~\cite{KNOLL_VOSKRESENSKY}. 
We substitute the resulting expression in 
eq. (\ref{EMISSIVITY}), carry out the neutrino phase-space integrals,
and sum over their three flavors. The final result for the 
emissivity is 
\be\label{E_NUNU}
\epsilon_{\nu\anu}=
\frac{2G^2}{5(2\pi)^7}(c_V^2+3 c_A^2)\, m^2 \, T^7\, {\cal I}(T),
\ee
where $T$ is the temperature,
\be \label{INT_NUNU}
{\cal I}(T) &=&
\int_{0}^{\infty}\!\! dz \, z^5\, g_B(z) 
\int_{-\infty}^{\infty}\!\! dy
\left[f_N(y)-f_N(y+z)\right] \int_{0}^{\infty}\!\! dx 
\frac{\chi(y,x)}{\left[y - \xi(x)\right]^2 +\chi(y,x)^2/4}
\nonumber \\
&&\Biggl\{ 
{\rm arctan}\left[\frac{\xi(x)+\zeta(z) - y - z + 
2 \sqrt{x\zeta(z)}}{\chi(y+z,x)/2}\right]\nonumber \\
&&\hspace{2cm}-{\rm arctan}\left[
\frac{\xi(x)+\zeta(z) - y - z - 2 \sqrt{x\zeta(z)}}{\chi(y+z,x)/2}
\right]\Biggr\}, 
\ee
with the quantities $\xi(x) = \ep_p/T$, $\chi(y,x)= 
\gamma(\omega,p)/T$,  $\zeta(y) = \ep_q/T$ 
being functions of  $x = \ep^0_p/T$, $y = \omega/T$, $z = \omega'/T$;
here $\ep_q$ is the recoil energy.
Note that eq. (\ref{E_NUNU}) contains only universal 
couplings\footnote{The values of parameters of 
weak interaction are $c_V = -1$, $c_A=-(F+D)$ for neutrons, 
$c_V = 1-4\, {\rm sin}^2~\theta_{W}$, $c_A=F+D$ for protons,  
$c_V = -2+4\, {\rm sin}^2~\theta_{W}$,  $c_A=-F$ for $\Sigma^-$
hyperons  and zero for $\Lambda$ hyperons; 
here $F=0.477\pm 0.012 $, $D=0.756\pm 0.011$ 
and ${\rm sin}^2~\theta_{W} = 0.23$,
where $\theta_W$ is the Weinberg angle. As noted above, 
the particle-hole vertex corrections renormalize these 
values of $c_V$ and $c_A$.}  and constants as a prefactor, 
all the nuclear many-body effects 
are absorbed in the integral ${\cal I}(T)$, 
which is valid at arbitrary densities and temperatures
(in particular does not rely on any type of 
low-temperature expansion). At low-temperature our result
can be simplified by expanding the 
spectral function (\ref{SF}) to the leading order in $\gamma$. 
Performing the same steps leading to eq. (\ref{E_NUNU}),  
we find in the first order $\gamma$
\be\label{E_NUNU_LOW}
\epsilon_{\nu\anu}^{(1)}=
\frac{4G^2}{5(2\pi)^6}(c_V^2+3 c_A^2)\, m^*p_F Z(p_F) T^7\,
\int_0^{\infty}\!\! dz\, z^7 \, g_B(z)\, \chi(z)
\frac{{\cal P}}{\left(z -\zeta(z)\right)^2},
\ee
where $p_F$ is the baryon Fermi momentum, 
$m^*$ is the effective mass, and
$Z(p_F)$  is the  wave-function renormalization; 
${\cal P}$ stands for the principal value
integration. In this limit
the width is essentially the reciprocal of the 
quasi-particle life-time in a Fermi-liquid, hence 
$\gamma\propto T^2 \left[1+(\omega/2\pi T)^2\right]$. Substituting
this  in  eq. (\ref{E_NUNU_LOW}) we recover the 
$T^8$ temperature dependence of the neutrino 
bremsstrahlung rate \cite{FRIMAN_MAXWELL,VOSKRESENSKY1}.
Note that according to ref. \cite{FRIMAN_MAXWELL}
the matrix element for the Fermi transition vanishes  
in the limit $q\to 0$ in Born approximation. 
This cancellation occurs whenever 
the off-shell source particle propagation between the strong and
weak interactions is odd under time-reversal,  which is true in the 
quasi-particle approximation neglecting the recoil term. We can 
not make an accessment on this cancellation, 
as the quantum interference diagrams are not included so far in our 
treatment. Should this cancellation indeed emerge as a consequence
of the current conservation requirements,\footnote{Note that 
the arguments of ref. \cite{FRIMAN_MAXWELL}, 
perfectly valid in the limit discussed there, are not directly 
applicable to our discussion, as neither the particle width is an 
odd function of the frequency, nor the recoil term is 
neglected.} our numerical expression
for the neutrino bremsstrahlung would overestimate the
rate by a small 
 factor $c_V^2/3c_A^2$.

A numerical estimate of the neutrino emissivity eq. (\ref{E_NUNU})
for the case of the neutron bremsstrahlung gives
\be\label{NUM_E_NUNU}
\epsilon_{\nu\anu} =
\left(7.56 \times 10^{18} ~{\rm erg ~cm}^{-3}~{\rm s}^{-1}\right)
 {T_9}^7\, {\cal I}(T).
\ee
where $T_9$ is the temperature in units $10^9$ K.
The triple integration in eq. (\ref{INT_NUNU}) is carried out
numerically after the functions $\xi, \chi$ and $\zeta$ 
are derived from the microscopic theory. 
Table 1 shows the results of the calculations of the 
integral ${\cal I}(T)$ and compares the neutrino emissivity,
eq. (\ref{NUM_E_NUNU}), with the result of 
Friman and Maxwell\cite{FRIMAN_MAXWELL}.
\begin{table}
\caption[]
\footnotesize{The upper number for a density-temperature pair 
is the value of the triple integral ${\cal I}$, the lower number
is the ratio of the eq. (\ref{NUM_E_NUNU}) to the result of Friman 
and Maxwell, eq.~(66a)}.
 
\begin{tabular}{ccccccc}
\\
\hline
 $T$ (MeV) & 0.5 & 2 & 5  & 10 & 20 & 30\\
\hline
$n_s$  &34.0  & 19.3 & 18.3 & 8.6   & 3.0  & 0.4\\
       &2.46 & 0.35 & 0.13 & $3.1\times 10^{-2}$
       &$5\times 10^{-3}$ & $5\times 10^{-4}$ \\
\hline
$2n_s$ &33.0 & 18.8 & 17.64 & 5.49 & 3.84 & 2.12\\
       &2.38 & 0.34& 0.127 & $3.3\times 10^{-3}$ 
& $ 7.0\times 10^{-3} $& $4.3\times 10^{-4}$ \\
\hline
\end{tabular}
\end{table}

\section{Discussion}

In this work we recovered the relation [eq. (\ref{EMISSIVITY})] 
between the neutrino emissivity and the polarization function 
of the nuclear medium from the Kadanoff-Baym 
transport theory. We derived the bremsstrahlung 
rates by computing the neutrino 
transport self-energies which are driven by neutrino
coupling to baryons with spectral width. The diagrammatic expansion 
for the polarization function of the medium is constructed 
with fully dressed particle propagators, as originally 
suggested by Knoll  and Voskresensky \cite{KNOLL_VOSKRESENSKY}. 
As a result the LPM effect, i.e. the influence of the multiple 
scattering during the radiation process, is included in the rate 
of neutrino  emission. We have kept only the 
leading order term in the expansion given by the  
one-loop polarization; more complicated 
diagrams are not considered at this stage.
Already at the leading order the rates are 
well behaved in the soft neutrino limit and exhibit 
LPM suppression when $\gamma/T \ge 1$.
Since this ratio at low temperatures
is proportional to $T$ the LPM suppression 
increases with increasing temperature;
in our case the typical scale for the onset of the 
LPM effect is  $T\sim 5$~MeV. A comparison to the result of 
Friman and Maxwell \cite{FRIMAN_MAXWELL} shows a strong suppression 
at relatively high temperatures and an enhancement by a factor 
of two in the low temperature limit.\footnote{Note that 
the extrapolation of the Friman and 
Maxwell's result to high temperatures 
is limited to the assumption that the quasi-particle picture
can still be applied and that the quasi-particles are 
lying close to their Fermi surface (the expansion parameter
being the ratio of the temperature over the Fermi energy).}
We believe that the latter low temperature enhancement
is mainly due to the differences 
in the  treatment of the nuclear interaction in the medium.
For densities at and above the nuclear saturation density 
effects beyond those included in the Brueckner formalism,
such as an enhancement of the strong interaction amplitude
due to the in-medium modifications of the pion degrees
of freedom, could compete with the LPM suppression, c.f.
\cite{VOSKRESENSKY1}.

Another feature of the expansion above is that 
the quantum interference terms
appear only in higher order diagrams.
The latter correspond e. g. to particle-hole 
insertions which transfer momentum between
the particle and the hole Green's functions  \cite{KNOLL_VOSKRESENSKY}.
It is well known that at $T=0$ inclusion of these interference 
terms is required to fulfill the Ward-Takahashi 
identities \cite{WAMBACH}. At finite temperatures
the interference corrections are reduced; an explicit evaluation
at $T\neq 0$ has to be performed yet.

This work has been supported by the Stichting voor Fundamenteel 
Onderzoek der Materie (FOM)
with financial support from the Nederlandse Organisatie voor
Wetenschappelijk Onderzoek (NWO).

\newpage
\end{document}